# Flares in Gamma Ray Bursts (II)

*Guido Chincarini[1,2], Raffaella Margutti[1,2], Jirong Mao[2], Francesco Pasotti[1,2], ,Cristiano Guidorzi[2], Stefano Covino[2], Paolo D'Avanzo[2]*

*& Swift Italian team*

1. *Università degli Studi di Milano – Bicocca. Piazza della Scienza, 3, 20126 Milano*
2. *OsservatorioAstronomico di Brera – INAF. Via E. Bianchi 46, 23807 Merate [LC]*

Abstract.

We illustrate some of the preliminary results obtained with a new sample of flares and a new analysis. In these proceedings we deal mainly with the analysis related to the flare energy and describe the work in progress to measure the average flare luminosity curve. We discuss in brief GRB050904 and GRB050724 for matters relevant to this work. In particular we measure the contribution given to the flares by GRB050904 and give a new interpretation for the decaying early XRT light curve of GRB050724. We briefly illustrate, for GRB050724, the first evidence that the early decay is given by the subsequent emission of events with Width/$T_{Peak}$ < 1 and the total energy of these events is larger than the energy emitted during the prompt emission spike showing, indeed, that not only the central engine may still be active after hundreds of seconds of the first spike but that this may still be part of the prompt emission.

Keywords:  Gamma ray source: Gamma ray burst.

1. Introduction.

In the work we presented at the GRBs Nanjing meeting [Astro-ph 0809.1026] we briefly outlined the characteristics of the new XRT flare sample we analyzed. We decided to use for the fit the function suggested by Norris et al., 2005, since it seemed to us a rather robust and convenient analytical function. It fits properly and very accurately single and multiple flares. In a few cases we fitted the light curves first using a simultaneous  broken power law plus Norris function fit (for the underlying light curve and the flares, respectively), while in most case the fits were done separately. Both procedures were perfectly acceptable giving very similar results within the errors.  In this way we were able to derive a rather robust estimate of the fluence of the flares using the standard definition of T90. In other words the fluence is computed as the integral of the profile defined by the T90 temporal interval.

If the central engine, whatever the mechanism might be, remains active or reactivates long after the prompt emission [as suggested by the flare activity] the total energy involved in the phenomenon is  a crucial parameter  since it will constrain the mechanism at work. A simple scenario expected after the collapse of a massive star, but also after the merging of two neutron stars, is that of a black hole surrounded by an accretion disk. Flares could be produced by accretion of matter after the breaking of the disk due to the setting up of various instabilities either gravitational or viscous. Another possibility may be the sudden increase of the accretion rate for very short period of time:  a mechanism capable of modulating the flow is not yet known, see Kumar et al., 2008 for an interesting model. Finally it is not clear whether the distribution of flares in time and energy is a completely stochastic event or it is correlated with the prompt emission energy or its frequency is a function of time. Flares energetics may be related to the mass of the central black hole and whether or not this is true can be investigated by a careful analysis of the observations, with particular attention to the energetic, and related modeling.

As we mentioned in the Nanjing meeting paper it is mandatory to estimate the probability of having a flare at a given time after the burst. In other words we need to know the global emission of GRBs as a function of time. Lazzati et al., 2008, approached as well this problem determining the average flare light curve using the data of the sample by Chincarini et al., 2008, Falcone et al., 2008 (and references therein). They conclude that the flare luminosity changes with time as $L \propto t^{-1.5}$. This correlation, derived from the above flare sample, would obviously help in understanding the mechanism at work and may even be capable of distinguishing between the black-hole accretion disk model and the spinning-down pulsar. The problem is, as we mentioned in the Nanjing paper, that we feel rather uncertain about the completeness of the sample because of the number of variables involved in the observations and in the analysis. Nevertheless the goal is certainly worth the effort to do the best we can in this direction. In this presentation we will basically focus on this matter and illustrate some preliminary results of a new sample, Chincarini et al., 2008, in preparation.

Note that the sample discussed here and in the Nanjing proceedings refers to early and reasonably bright flares. The complete data analysis with the addition of the late flare sample that we have already analyzed will be given elsewhere. We used a $H_o$ = 70 km/s/Mpc, $\Omega_\Lambda$ = 0.7, $\Omega_m$ = 0.3 cosmology.

2. The Energy emitted during flaring.

In this new sample we have 20 GRBs with redshift: of these 11 have a single measured flare, 7 have two flares and 2 show three flares; for a total of 31 flares. This sample refers primarily to early and well defined flares (the profile could be clearly fit): we call this the basic sample. It does not include, for instance, the short GRB050724 (important for the late flares it shows); GRB050820A since we miss the observations of the flare decaying part; GRB050904 since it is rather difficult to fit the late flares with the selected profile; GRB 060124 because the function is also difficult to fit in a simple way and likely it is one of those cases where the XRT flare may be identified with the prompt emission. All these cases are however dealt with separately and added to the sample when needed, see section 3.

The distribution in redshift and energy of the basic flares sample is illustrated in Figure 1.

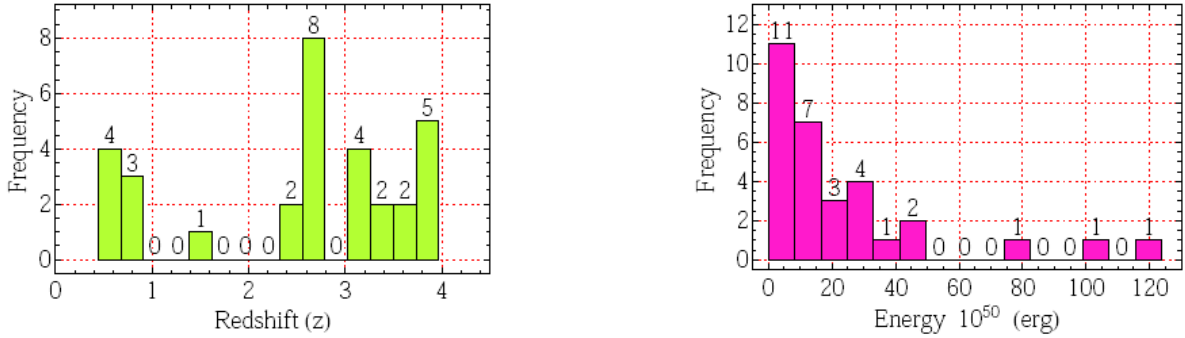

*Figure 1. Left: Distribution in redshift of the basic sample of 31 flares. Right: Distribution in energy (isotropic) for the 31 flares of the basic sample.*

The energy of flares when plotted versus redshift does not show a correlation. At low redshifts, however, very bright flares are missing while at high z we have a rather large spread in energy likely due to a volume effect.

As it was illustrated in the Nanjing proceedings the minimum XRT fluence we measured in this sample is about $10^{-8}$ – $10^{-8.5}$ erg cm$^{-2}$. To some extent this reflects in the plot on the right of Figure 2 where we plotted the ratio XRT flare energy / prompt emission energy versus the energy observed in the prompt emission. On the left panel we plot the XRT flare energy [0.3 – 10 keV] versus the prompt emission energy [15 – 150 keV]. In the high energy region of the plot the points are rather random: the effect of an observational bias is currently under investigation. We want to call attention to the low energy flares: one of these is a short. While at present no robust conclusion can be drawn, the observations indicate first that the energy we have in the afterglow is generally less than the energy we have in the prompt emission; second, the energy we have in flares is generally less or equal than the energy we have in the afterglow. Therefore the energy in flares is somewhat related, in a broad sense and especially including short bursts, to the energy observed during the prompt emission.

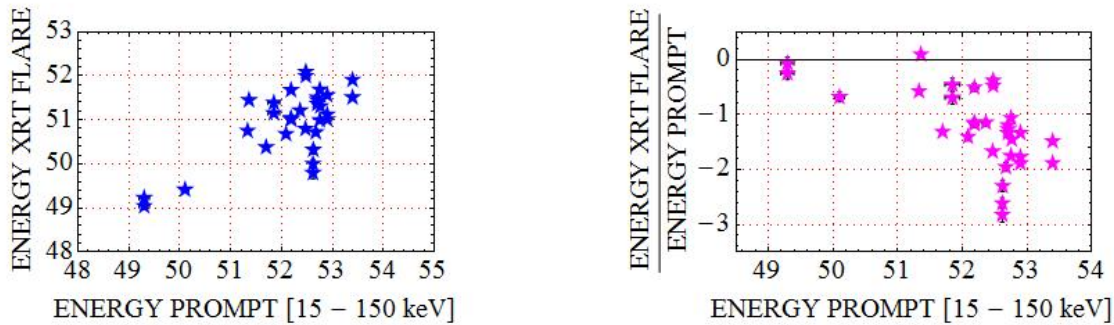

*Figure 2. Left panel: Energy measured in each single flare of the sample versus the prompt emission energy of the GRB. The three points in the bottom left part of the diagram refer to two flares detected in the short GRB070124 (Log [Energy Prompt] $\approx$ 49.2) and one flare associated with GRB060512 ( Log [Energy Prompt] = 50.1). Right panel:*

*The same data plotted in a different form to evidence, also in relation to Figure 8 of the Nanjing proceedings, eventual selection effects due to the limited sensitivity of the sample.*

Figure 3 is the same as Figure 2 left except that for each GRB we summed the energy emitted by the single flares..

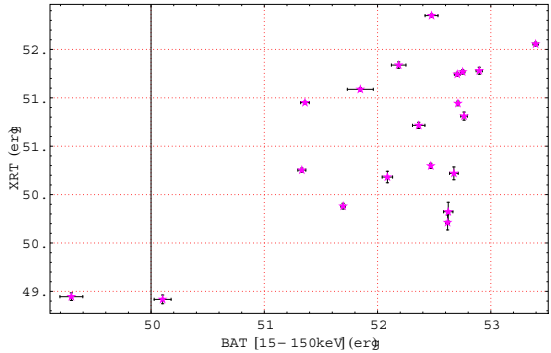

*Figure 3. Same as in Figure 2 with the difference that for each GRB we added the contribution of the flares that occurred in that GRB.*

3. Flare occurrence distribution.

As we mentioned in the introduction a function to be estimated is the probability of flare occurrence or flare luminosity as a function of time accounting for all the flares observed.

We use here the simple method of estimating the flare energy and derive the flare mean luminosity by dividing by the T90 in the rest frame. In other words we approximate using a rectangular profile of the flare and conserving the energy [see however later for the approach we are working on]. Among the flares that are not included in the basic sample we have GRB050820A and GRB060124 that needed to be handled separately and that may also cause large uncertainties: for GRB050820A only the rising part of the profile has been observed: the giant flaring activity of GRB060124 is likely part of the prompt emission (see Romano et al. 2006), the fit with the adopted profile is not possible [using a multiple function may do it however].

Accounting, in addition of the flares basic sample, for GRB050820A we increase the early activity. Accounting for GRB060124 we add a large contribution to the average flare luminosity between 140 and 240 s (Figure 4 right). In these plots we did not add the contribution of GRB050904 since we limited ourselves to the early flares (the first flare of GRB050904 increases only slightly the maximum at 60 s rest frame but is important for the late part of the curve that we discuss elsewhere).

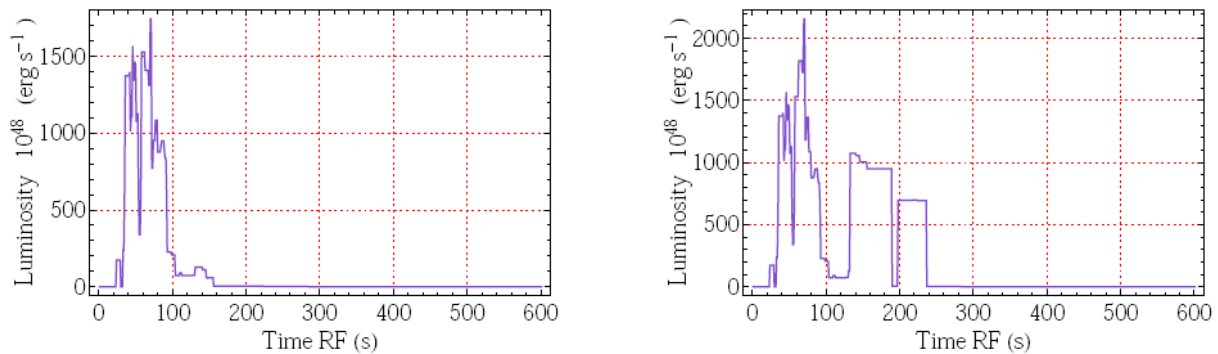

*Figure 4. left: Plot of the distribution of the flare activity summing up the 31 early flares we considered. Right: Adding to the previous sample GRB050820A, GRB060124.*

GRB050904 is the farthest GRB detected by Swift with z = 6.29. As illustrated in Figure 5 there is a bit of subjectivity in deciding the underlying light curve. The one we selected is rather common for the GRBs detected by Swift and in any case minor differences in this curve does not affect the parameters of the flare activity.

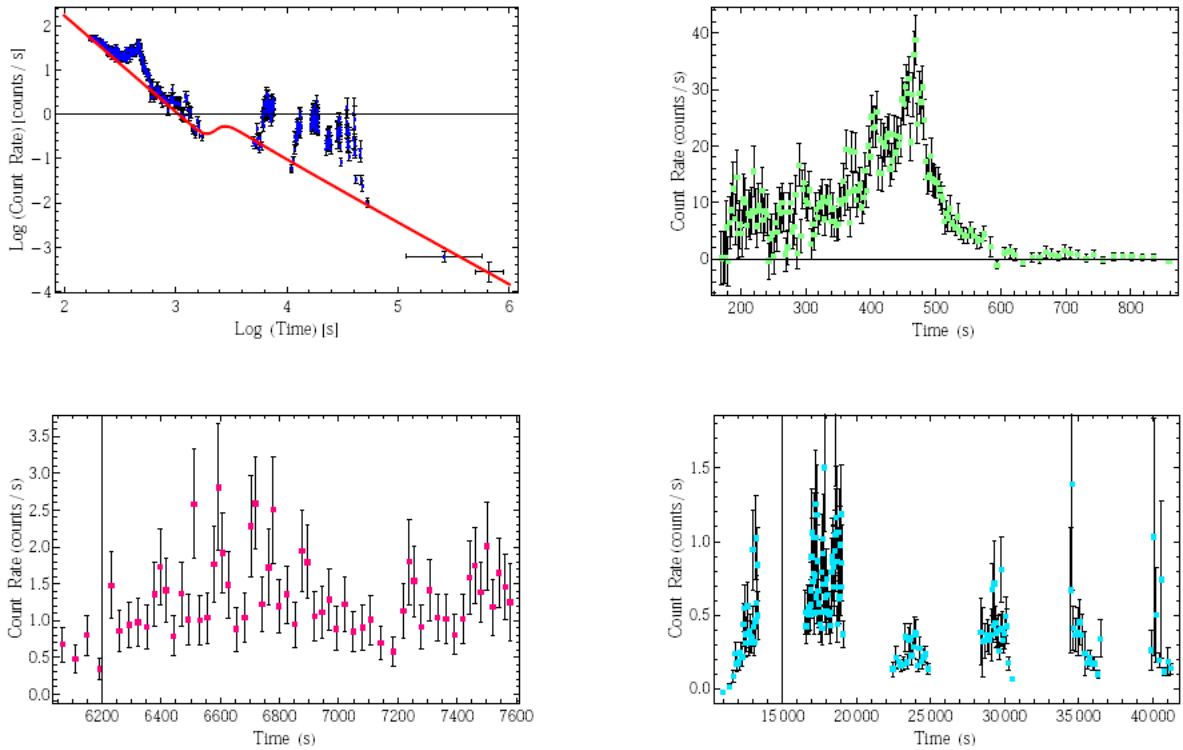

*Figure 5. Top left: XRT observations and broken power law fitting the data selected for the underlying light curve. Top Right: First flare after subtraction of the underlying continuum, Bottom left: Second flare, Bottom right: late sequence of flares. The assumption is that the activity keeps going and a functional fit would be very uncertain.*

The first flare [in the following all the time values are quoted in the frame of the source], this is easy to single out, is the more energetic (1.6 $10^{52}$ erg) and lasts about 87 seconds, Figure 5 top right. The second flare (from 700 to 1182 s) is incomplete, Figure 5 bottom left. The rest of the flare activity, from about 1493 to about 5492 s is likely related to interrupted activity and may be considered equivalent, in the framework of Figure 4, to a single flare, Figure 5 bottom right. The energy computed in the rest frame light curve under these assumption is respectively 7.61 $10^{51}$ erg for the second flare and 3.27 $10^{52}$ for the third flare.

The first flare is rather well fitted by a blend of 4 profiles and the related energy can be computed quite easily. For a single flare it is possible to compute the luminosity decay (see also Lazzati et al., 2008) in various ways. Based on the new analysis, we plot in Figure 6 the counts as a function of time and that clearly follow a power law. Note however that even a visual inspection of the large flares in GRB051117A shows that this is not the case for all GRBs. Even if in this sample we do not deal with late flares, these seem to be rarer than expected, so that the new larger sample may have characteristics that differ somewhat from the previous sample by Chincarini and Falcone, (all evidence is for a lower mean luminosity for the late flares).

The fit, and therefore the early flare, of the short burst GRB050724 has always been somewhat uncertain due to its smooth initial decline. The BAT 15 – 25 keV light curve shows a little weak and broad bump at about 100 s overlapping in time with the XRT early observations. This calls also for morphology problem, and related physical interpretation, of which a classical example is GRB060616, (Della Valle et al., (2006), Gehrels et al., (2006)).

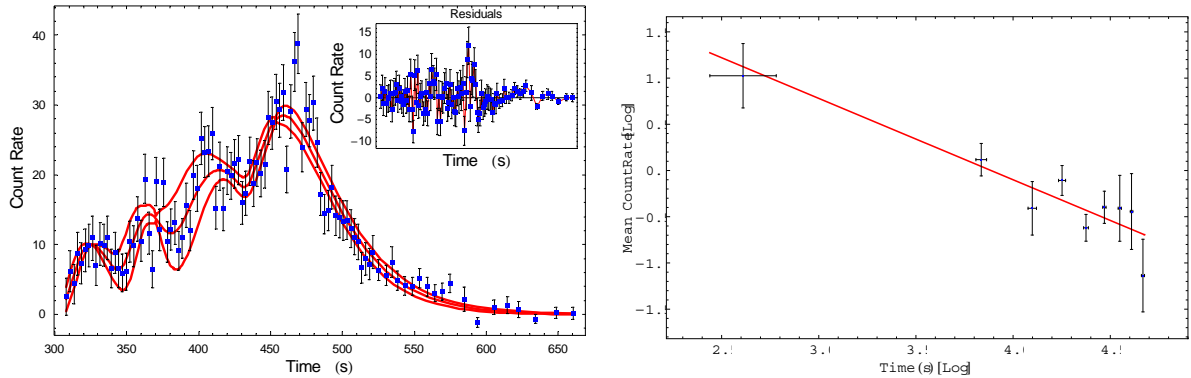

*Figure 6. Left: Fit of the first flare observed in GRB050904 using four functions and taking into account the errors. Right: the average counts as a function of time for all the flaring activity observed in GRB050904.*

To better understand this light curve we derived it both binning at fixed number of counts interval and binning according to a constant S/N. The matter became finally clear and is illustrated in Figure 7. After the very brief and intense spike observed with the BAT instrument that classifies it as a short GRB, we have a soft faint and broad bump (15 – 25 keV) starting at about 40 s (observer time). This bump is very strong in the XRT observations starting at about 80 seconds after the trigger time. The whole early light curve is nicely fitted, $\chi^2$ / dof = 0.92, by a sequence of 4 events. In Figure 7 we show the light curve and the fit of the first 225 seconds. For clarity in the figure we plotted the data with constant number of counts, however the fit was based on the constant S/N light curves, total of 768 data points. The fitting of the late light curve, and flare profile, remains unchanged.

Let's remind the goal of the analysis. The model is that of a more or less standard afterglow light curve superimposed to which we have variability of various intensity and time scales. We refer to the most intense superimposed emissions as flares. These flares show that the activity of the central engine lasts for periods that are much longer than the duration of the prompt emission. It seems furthermore that flares may occur at any time. On the other hand it is fundamental to understand whether the luminosity changes statistically as a function of time and whether the probability of occurrence is a function of time using a large and reasonably controlled sample.

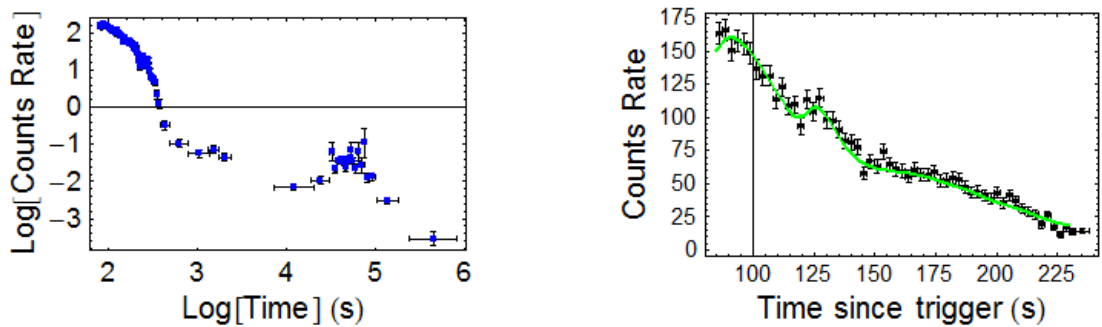

*Figure 7. Left: XRT light curve of GRB050724 with constant binning, Right: Fit of the first 225 seconds. Note however that the fit was carried out on the constant S/N binning light curve (625 data points for this part).*

A first indication of time dependence was given by Chincarini et al. 2007. In that work we pointed out that the peak intensity of the flares was decreasing with time with no big change in the energy (since late flares last much longer). This is an indication of decreasing mean luminosity. On the other hand in addition to the easily visible and measurable flares we notice that: a) it is often impossible to measure the flare fluence; b) quite often even the rising part of the "flare" is dominated by merging of many flares (see Figure 5 and 6 or consider the long lasting flare activity of GRB051117), c) we often observe mini-flare and low intensity scale variability that while contributing little to the general picture may be important in characterizing the variability. Finally, given the distance of some of the GRBs (the farthest one is at z = 6.29), we must account for the effect of the redshift on the pass band since these changes become important. On this GRB the rest frame XRT pass band is between 2.19 keV and 72.9 keV so that the luminosity is largely overestimated (and it is a function of the spectral index) compared to a GRB at z = 0.4 for instance. It is necessary, therefore, to compare luminosities in the same rest-frame pass band in order to maintain homogeneity. Even using solely our basic sample, for instance, the GRBs we consider are between a minimum redshift z = 0.54 (GRB060729) and a maximum z=4.9 (GRB060510B) with a considerable change in the passband.

The analysis described above is close to completion and ready for publication. On the other hand working on this sample, we realized that we should also take an alternative approach. And since in this context we wanted to discuss the data and the procedures, we briefly describe the new approach by which we started to measure the average flare luminosity curve and apply it to a small selected sample of bright flares.

To this end we studied the 13 brightest long GRB X-Ray afterglows detected by Swift XRT up to March 2008. We require these events to show a prominent flaring activity and have a measured redshift. This sample comprises: GRB050730, GRB060124, GRB060210, GRB060418, GRB060510B, GRB060526, GRB060607A, GRB060729, GRB060814, GRB060904B, GRB061121, GRB071031, GRB080310. Light curves and spectra were extracted in a standard way using the latest Heasoft package available at the time of the analysis (see the upcoming paper for details).

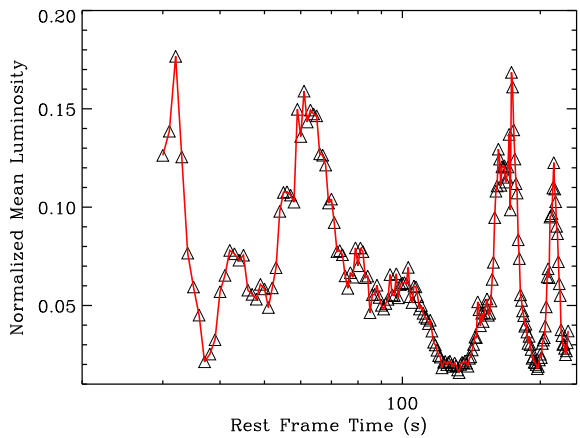

*Figure 7. Normalized average flare luminosity light curve obtained as explained in the text. A common rest frame energy band (1.77-15.4 keV) has been used.*

The spectral information was used to extract the light curves in a common energy band (1.77-15.4 keV).

We concentrated on the very first part (t<1000 s) of these events (where the vast majority of X-Ray flares are detected), and fit their light curves with broken power laws: the portions of the light curves dominated by flares were excluded from the fitting procedure (see figure 5 as an example). Despite the somewhat subjective approach, this procedure does not assume any a-priori flare shape: the flaring activity of each burst is simply obtained as the difference between the detected signal and the best fit broken power law modeling the underlying continuum.

We then normalized the flaring activity of each burst dividing the previous result by its maximum value. The average flare luminosity curve is finally obtained as the sum of the normalized flaring activity of each burst calculated in the same rest frame energy band normalized by the number of events showing emission in the interval of time considered. The result is shown in Figure 7. This is a very preliminary plot, given the small size of the sample. Nevertheless it is possible to note that no clear sign of a declining flaring activity with time is apparent in the first ≈200 s (rest frame value) of X-Ray emission

4. Summary & Conclusions

Preliminary results on the new sample of flares show that the Norris 2005 profile (but equally well seems to fit the profile by Kocevski et al. 2003) fits quite nicely not only the prompt emission spikes but also the flares observed during the afterglow by the XRT instrument. In this work we focused mainly on the procedure and gave some indication on the energy emitted in flares. We find an indication that GRBs where the prompt emission energy is very large have in general also rather energetic flares while burst with rather faint prompt emission tend also to have low energy flares and this seems to be true when comparing especially long and short GRBs. However selection effects may be very strong and may need to be tracked down by simulations.

We briefly touch upon the new analysis we carried out on GRB050724. This new analysis shows that the early XRT emission, detected also in the soft band of the BAT instrument, is due to a sequence of at least four events with $\Delta t/t < 1$ (mean value 0.43 with dispersion 0.21). The BAT prompt emission energy (15 – 150 keV) of the short spike is of about $1.77 \times 10^{50}$ erg while in the interval 85 – 352 s since trigger the XRT (0.3 – 10 keV) emission of the 4 events is about $1.85 \times 10^{50}$ erg [error about 5%] and it is this part of the light curve that generates the steep early slope. It is quite likely we are still dealing with an active engine. This pose more seriously the question of the flare morphology also in relation to the debate related to the observations of GRB060614. We will expand elsewhere on this matter.

The origin of flares remains unsolved. The main problems we need to solve observationally, and the work is in progress, are related to the frequency and probability of occurrence as a function of time. The average luminosity light curve will be also one of the products of this analysis. From the theoretical point of view we need a detailed mechanism and the understanding of the energetic. If we have gravitational instabilities and accretion we should also find out whether the emission is jetted or not. But assuming we have equatorial accretion do we expect a jet?

Acknowledgments:

The GRB Research is being supported by ASI contract SWIFT I/011/07/0 and by MIUR PRIN 2007TNYZXL. The continuous support by the Università degli Studi di Milano Bicocca is also acknowledged.